\documentclass{article}

\usepackage{amsmath,amssymb}
\usepackage{graphicx}
\usepackage{dcolumn}
\usepackage{bm}

\begin{document}

\title{Nonparametric estimation of the heterogeneity of a random medium using Compound Poisson Process modeling of wave multiple scattering.}

\author{Nicolas Le Bihan$^{(1)}$ \& Ludovic Margerin$^{(2)}$\\
(1): GIPSA-Lab, Dpt. Images et Signal, \\
961 Rue de la Houille Blanche,\\
Domaine Universitaire,\\
38402 Saint Martin d'H\`eres Cedex, France \\
 \\
(2): CEREGE,\\
Europ\^ole M\'editerran\'een de l'Arbois,\\
BP 80,\\
13545 Aix En Provence cedex 04, France}%

\date{\today}

\begin{abstract}
In this paper, we present a nonparametric  method to estimate the heterogeneity of a random medium from the  angular distribution of intensity transmitted through a slab of random material. Our approach is based on the modeling of forward multiple scattering using Compound Poisson Processes on compact Lie groups.
The estimation technique is validated through numerical simulations based on radiative
transfer theory.
\end{abstract}


\maketitle

\section{Introduction}

Small-angle scattering  of electromagnetic, acoustic or elastic waves constitutes the basic tool to characterize  porous media  \cite{porous1999}, solid suspensions \cite{xu2002}, or  Earth's lithosphere \cite{sato98}, to cite a few examples only. In many cases, the probing wave is assumed to undergo single scattering only and multiple scattering is often considered as a nuisance. The main reason
 is that  in the single-scattering regime, the diffraction pattern of intensity, i.e. the {\it phase function}
 is  related to the power spectrum  of the medium
 heterogeneities in a simple manner \cite{rytov1989}.
In the field of optics, some solutions  have been proposed to invert for the size of particles 
in the multiple-scattering regime \cite{hirleman1991}, based on various  small-angle approximations
of the radiative transfer equation \cite{ishimaru1978}.   In this work, we develop an alternative approach based 
on the  Compound Poisson Process   (CPP) model of  wave multiple scattering  in random media.  
Our model is generic and makes little assumption on the  underlying wave equation.
The random medium is described by  its heterogeneity power spectrum which refers to the Fourier transform of the spatial 
two-point correlation function. This enables us to treat 
on the same footing turbid media like Earth's crust or the atmosphere, aggregates of particules, and
 porous media \cite{torquato2002}. The finer details of the microstructure encapsulated in higher
moments  of the random field $(>2)$ are not considered in our approach. 

The organization of the paper is as follows: In section II, the necessary mathematical
background on non-commutative harmonic analysis is provided.
 In section III, we develop the Poisson process approach to multiple scattering.
 In section IV, we  validate the model through comparisons with Monte Carlo
simulations of the radiative transfer equation.  
The inverse problem is examined in section V, where we propose an estimator of the power spectrum
of heterogeneities  and discuss its statistical properties.
In section VI, the inverse method is validated  for realistic examples of random media through numerical 
simulations. We consider waves propagating
in a turbid medium described by the Von Karman correlation function, which is commonly used 
to  represent  geophysical media. We demonstrate the possibility to estimate 
the power spectrum of the random fluctuations from the angular distribution of intensity
transmitted through a slab of material.  We find that the CPP model is effective before the diffusive regime sets in, i.e.
when the   slab thickness is less than one transport mean free path.


\section{Noncommutative Harmonic Analysis}
\label{Section_NCHA}
In this section, we summarize noncommutative harmonic analysis results that are important for the present work. We are interested in the harmonic analysis of probability density functions ({\em pdf}) over compact Lie groups, in order to develop a nonparametric approach for the estimation of their characteristic functions \cite{Gre63}, {\em i.e.} their Fourier transform, in Section \ref{Section_estimation}.

Consider the direction of propagation of a particule or a wave  represented by the normalized vector $\mu \in {\mathcal S}^2$, where ${\cal S}^2$ is the unit sphere in ${\mathbb R}^3$. We denote by $\mu_N$ the direction of propagation of the particule after $N$ scattering events. Such scattering events are considered random so that $\mu_N$ is a random variable on ${\cal S}^2$. The relation between the direction after $N-1$ scattering events, {\em i.e.} $\mu_{N-1}$, and the direction after $N$ scattering events $\mu_N$ can be written as: 
\begin{equation}
\mu_N={\bf r}_N\mu_{N-1} ,
\end{equation} 
where ${\bf r}_N$ is a random element of the rotation group $SO(3)$. This follows from the transitive action of $SO(3)$ on ${\cal S}^2$. Assuming that a particule enters the random medium with initial direction of propagation $\mu_0$, its direction of propagation after $N$ scattering events in the medium is given by:
\begin{equation}
\mu_N={\bf r}_N{\bf r}_{N-1}\ldots{\bf r}_2{\bf r}_1\mu_0=\prod_{n=1}^N{\bf r}_n\mu_0,
\label{mun_prod}
\end{equation}
where each ${\bf r}_n$ represents the random action of the $n^{th}$ scatterer, {\em i.e.} the ${\bf r}_n$ are $SO(3)$-valued random variables. 
In the sequel, we present some results about the {\em pdf} of random variables of the form (\ref{mun_prod}). We make use of known results in harmonic analysis on compact Lie groups \cite{Dieu80,Bro85,Sal08}.

\subsection{Harmonic analysis on $SO(3)$ and ${\mathcal S}^2$}
\label{Subsection_HASO3}
Let us denote by ${\cal L}^2(SO(3),d{\bf r})$ the space of square integrable functions on $SO(3)$ with respect to the Haar measure $d{\bf r}$ \cite{Dieu80}. A probability density function ({\em pdf}) on $SO(3)$ is a function $f \in {\cal L}^2(SO(3),d{\bf r})$ which obeys the constraint $\int_{SO(3)}f({\bf r})d{\bf r}=1$. A {\em pdf} on $SO(3)$ can be decomposed over the complete orthonormal basis of Wigner-D functions $D_{l}^{pq}({\bf r})$, with ${\bf r} \in SO(3)$, $l \in {\mathbb Z}^+$, $p,q \in {\mathbb Z}$. This means that given a random variable ${\bf r} \in SO(3)$, its {\em pdf} $f(\bf r)$, can be expressed as the infinite series:
\begin{equation}   
  f({\bf r})=\sum_{l \geq 0}\sum_{p,q=-l}^l(2l+1)\hat{f}_l^{pq}\overline{D_{l}^{pq}({\bf r})},
\label{decomppdf}
\end{equation}
where: 
\begin{equation}
\hat{f}_l^{pq}=\int_{SO(3)}f({\bf r})D_{l}^{pq}({\bf r}) d{\bf r},
\end{equation}
and the overbar  means complex conjugation.
The set of coefficients $\hat{f}_l^{pq}$ are sometimes called the ``Fourier transform" of $f(\bf r)$ \cite{Roc08}. Note that it is possible to use a matrix notation, namely $\hat{{\bf f}}_l$, in which case, the elements of the $(2l+1)\times(2l+1)$ matrix $\hat{\bf f}_l$ are: $\left\{\hat{{\bf f}}_l\right\}_{pq}=\hat{f}_l^{pq}$. 

Similarly, let us denote by ${\cal L}^2({\cal S}^2,d\mu)$ the space of square integrable functions on the unit sphere ${\cal S}^2$ with respect to the invariant measure on the sphere $d\mu$. A {\em pdf} $w \in {\cal L}^2({\cal S}^2,d\mu)$ also satisfies: $\int_{{\cal S}^2}w(\mu)d\mu=1$. Similar to Equation (\ref{decomppdf}), a {\em pdf} $w$ on ${\cal S}^2$ can  be decomposed into an infinite series, using the well-known spherical harmonics $Y_l^p(\mu)$ with $\mu \in {\cal S}^2$, $l \in {\mathbb Z}^+$, $p \in {\mathbb Z}$. Thus, given a random variable $\mu \in {\mathcal S}^2$, its {\em pdf} $w(\mu)$ can be written as:
\begin{equation} 
w(\mu)=\sum_{l \geq 0}\sum_{p=-l}^l(2l+1)\hat{w}^p_l\overline{Y_l^p(\mu)},
\end{equation}
with: 
\begin{equation}
\hat{w}_l^p=\int_{{\mathcal S}^2}w(\mu)Y_l^p(\mu)d\mu .
\end{equation}
In this case, it is also possible to use a vector representation, where the elements of the $(2l+1)$ vectors $\hat{{\bf w}}_l$ are  $\hat{w}_l^p$ ($p=-l,\ldots,0,\ldots,l$). Again, $\hat{w}_l^p$ is called the Fourier transform of $w(\mu)$.

We now make use of  the following  fundamental property: the action of $SO(3)$ on ${\mathcal S}^2$ is transitive \cite{Bro85}. As a consequence, for any ${\bf r} \in SO(3)$ and $\mu \in {\mathcal S}^2$,  we have ${\bf r}\mu \in {\mathcal S}^2$. This Lie group action of the rotation group on the unit sphere implies that, given two {\em pdf}s $f({\bf r})$ and $w(\mu)$ taking respectively values on $SO(3)$ and ${\mathcal S}^2$, their convolution is a {\em pdf} over ${\mathcal S}^2$ and reads:
\begin{equation}
h(\mu)=\left(f*w\right)(\mu)=\int_{SO(3)}f({\bf r})w({\bf r}^{-1}\mu)d{\bf r},
\end{equation}
for any $\mu \in {\mathcal S}^2$. A very important and interesting consequence of the Peter-Weyl theorem \cite{Dieu80} is that this convolution equation is transformed into a multiplication in the ``frequency domain'' as follows:
\begin{equation}
 h(\mu)=\displaystyle{\sum_{l \geq 0}\sum_{p=-l}^l} (2l+1) \hat{h}_l^p\overline{Y_l^p(\mu)}.
\end{equation}
Using the matrix/vector notation introduced previously, we obtain the following relation: 
\begin{equation}\label{prod_Fcoeff}
\hat{{\bf h}}_l=\hat{{\bf f}}_l\hat{{\bf w}}_l ~ ~ \forall l ,
\end{equation}
which is a matrix-vector multiplication taking place for each degree $l$. As explained in \cite{Sal08}, two succesive rotations consist in an other rotation  ${\bf r}={\bf r}_2{\bf r}_1$. Moreover the {\em pdf} of ${\bf r}$ is the convolution of the {\em pdf}s of ${\bf r}_1$ and ${\bf r}_2$, namely $f_1({\bf r}_1)$ and $f_2({\bf r}_2)$, where convolution is again defined with respect to the group action: 
\begin{equation} 
\label{convol}
f({\bf r})=\int_{SO(3)}f_2({\bf t})f_1({\bf t}^{-1}{\bf r})d{\bf t}.
\end{equation}
In equation (\ref{convol}), ${\bf t}$ denotes an element of $SO(3)$, and $d{\bf t}$ stands for the Haar measure. Consequently, an $N-times$ product of independent random elements of $SO(3)$, {\em i.e.} $N$ consecutive random rotations, consists in a random rotation whose {\em pdf} is the $N-times$ convolution of the {\em pdf}s of the elementary rotations. Thus, if we denote by ${\bf r}={\bf r}_N{\bf r}_{N-1} \ldots {\bf r}_2{\bf r}_1$ the 
outcome of $N$ successive rotations, the Fourier transform of ${\bf r}$ can be expressed as follows:
\begin{equation}\label{eq_FT_prodSO3}
\hat{{\bf f}}_l=\displaystyle{\prod_{n=1}^{N}\left(\hat{{\bf f}}_n\right)_l} ~ ~ \forall l.
\end{equation}
which is a simple product of  matrices. 
 In (\ref{eq_FT_prodSO3}), $\left(\hat{{\bf f}}_n\right)_l$ denotes the matrix of Fourier coefficients of degree $l$ of $f_n$, the {\em pdf} of ${\bf r}_n$. 

\subsection{Symmetries}
\label{Subsection_Sym}

Using Eq. (\ref{prod_Fcoeff}) and (\ref{eq_FT_prodSO3}), it is possible to obtain the Fourier coefficients of any distribution on ${\cal S}^2$ that was multiply convolved by {\em pdf}s on $SO(3)$. In what follows, we will derive more specific results for distributions on ${\cal S}^2$ which are  functions of $\mu=\cos\theta$ only, with $\theta$ the scattering angle. Such functions are called {\em zonal} functions on ${\cal S}^2$  \cite{Dieu80}. This simplification relies on the property of statistical isotropy of the random medium. 
 As a consequence, the Fourier transform over ${\cal S}^2$  reduces to a Legendre expansion over $[-1;1]$ for zonal functions. 
 In anisometric random media where the correlation length depends on space direction, this simplification is not permitted.

In order to obtain simple results, we further impose that the initial distribution of directions of propagation $\mu_0$
 is  a zonal function. Such an assumption is not too restrictive and covers for example the case of an incident plane wave,
 i.e. $\mu_0$ is a Dirac distribution.
Using the ${\cal XZX}$ parametrization of $SO(3)$ \cite{Roc08},  the symmetry of the scattering process around the direction of propagation
 implies that  the {\em pdf} $w_{\mu}(\mu)$ of the direction of propagation $\mu$, parametrized using co-elevation $\theta$ (with respect to direction of propagation) and azimuth $\varphi$, is invariant with respect to $\varphi$. This allows us to expand $w_{\mu}(\mu)$ as follows:
\begin{equation}  
w_{\mu}(\mu)=\displaystyle{\sum_{l \geq 0}(2l+1)\hat{w}_l^0\overline{Y_l^0(\mu)}=\sum_{l \geq 0}(2l+1)\hat{w}_lP_l(\cos\theta)},
\label{EqS2_symm}
\end{equation}
where the $\hat{w}_l$ are scalar quantities. The rotational symmetry of the scattering process also implies that the {\em pdf} $f({\bf r})$ 
of random elements of  $SO(3)$  must be bi-invariant, i.e. invariant by left and right action on $SO(3)$  (see \cite{Dieu80} for details).
 Such a bi-invariant {\em pdf}  $f({\bf r})$ depends on   a single variable $\mu=\cos\theta$, and can therefore be expanded as:
\begin{equation}
f({\bf r})=\displaystyle{\sum_{l \geq 0}}(2l+1)\hat{f}_l^{00}\overline{D_l^{00}({\bf r})}=\sum_{l \geq 0}(2l+1)\hat{f}_lP_l(\cos\theta).
\end{equation}
 Considering the successive action of  $N$ {\em i.i.d.}  random rotations on the initial distribution of propagation directions $\mu_0$ 
leads to the following expression of the {\em pdf} $h(\mu_N)$ of $\mu_N={\bf r}_N{\bf r}_{N-1}\ldots{\bf r}_2{\bf r}_1 \mu_0$:




\begin{equation}
h(\mu_N)=\displaystyle{\sum_{l \geq 0}}(2l+1)(\hat{f}_l)^N\hat{w}_l P_l(\mu).
\label{coeff_convol_SO3S2}
\end{equation}
In the special case of an incident plane wave or a highly collimated beam, the initial direction obeys the following simple probability
distribution $w(\mu_0)=\delta(\mu - \mu_z)$ where $\mu_z$ is the vector pointing in the direction of propagation.
Introducing  this expression of $w$ in Equation (\ref{EqS2_symm}) yields:
$
\hat{w}_l=P_l(\mu_z).
$
By symmetry $\mu_z$  must coincide with the   co-latitude 0, {\em i.e.} the north pole $\theta=0$, which implies $\hat{w}_l=1$, $\forall l$. 
In such a case, eq. (\ref{coeff_convol_SO3S2}) becomes:
\begin{equation}
f(\mu_N)=\displaystyle{\sum_{l \geq 0}}(2l+1)(\hat{f}_l)^N P_l(\mu)
\label{dist_RandScatt_N}
\end{equation}
The {\em pdf} of $\mu_N$ possesses a simple Legendre expansion with  coefficients equal  
to the $N$th power of the Legendre coefficients $\hat{f}_l$ of the  $SO(3)$ random variables. 

Note that Equation \ref{dist_RandScatt_N} could have been obtained using the well-known summation formula for Jacobi polynomials (see Chapter 2 in \cite{Dieu80}) in the case of {\em zonal} functions on the sphere. The  approach developed in this section is more general because it could be applied to
 less symmetrical distributions, thereby allowing the modeling of more complicated scattering processes.

\section{The Compound Poisson Process model for multiple scattering}
\label{Section_cpp}

In this section, we develop the  multiple scattering model. We begin with a summary of useful results about Compound Poisson Processes (CPP) on compact Lie groups \cite{IMA08}. Note that we use the term $pdf$  when we refer to the function describing the angular
 pattern of scattering.   In Section \ref{Section_val}, we will substitute it with  {\em phase function} which is the usual terminology
 in the field of scattering in  random media.
\subsection{The model}
The implementation of the CPP model for the study of multiple scattering of particules/waves has already been described in \cite{PhysRevE.52.5621}. In \cite{PhysRevE.52.5621}, the authors were mostly interested in formulating the direct problem, {\em i.e.} in the way a CPP can be used to predict the output distribution of scattering angles in a scattering experiment. They demonstrated the ability of the CPP to model the  multiple scattering of electrons
and used some recursive integral equations to obtain the {\em pdf} of the CPP. The  CPP model introduced in  \cite{PhysRevE.52.5621} is based on 
the cumulative scattering angle, which is a  real valued random process defined as  a sum of real-valued random variables.
 In this work, we introduce a  CPP on a compact Lie group, namely the rotation group $SO(3)$. 
Our approach is more general and does  not rely on an {\em a priori}   small-angle approximation.

Let us consider a particule/wave entering a random medium at time $t=0$, with initial direction of propagation $\mu_0 \in {\cal S}^2$,
 as  illustrated in  Figure \ref{slab}. For the moment we neglect the possibility that the wave may escape the random
medium.  We consider a slab of random  material with mean free time $\tau$ and  velocity $c=1$ for simplicity.
 The normalization of the velocity implies $\lambda=1/\ell$ where $\ell$ is the mean free path.   
We seek to model the time evolution of the direction of propagation $\mu(t)$. 
Denoting by $N(t)$  the random number of scattering events that have occurred
after propagation during  time $t$ in  the random medium, the direction of propagation $\mu(t)$ can be written as:
\begin{equation}
\mu(t)=\prod_{n=0}^{N(t)}{\bf r}_n\mu_0.
\label{modelCPP}
\end{equation}
In order to model wave  scattering,  $N(t)$ is chosen as  a Poisson process, and is independent of the ${\bf r}_n$. 
To be fully consistent, we must impose ${\bf r}_0 = \mathbf{I}$, where $\mathbf{I}$ denotes the identity in $SO(3)$. This simply means
that the unscattered energy propagates in the direction imposed by the source.  
Clearly, the  ${\bf r}_n$, $n \geq 1$ are independent and identically distributed random rotations described by the {\em pdf} $f({\bf r})$. 
Our choice for  $N(t)$  implies the usual exponential distribution of times of flight between two scattering events,
 with parameter $\lambda=1/\tau$. 

\subsection{Probability density function of the CPP}
\label{Subsection_pdfCPP}
To complete our task, we need to relate the {\em pdf} of $\mu(t)$, denoted by $p(\mu(t))$, to the scattering properties of the medium.  
Keeping in mind the {\em i.i.d.} assumption for ${\bf r}_n$ and using conditional probability decomposition
  with respect to the Poisson process $N(t)$, one gets:
\begin{equation}
p(\mu(t))=\sum_{n=0}^{\infty}e^{-\lambda t} \frac{(\lambda t)^n}{n!}f^{\otimes n}({\bf r})*w(\mu_0),
\label{cpp_pdf}
\end{equation}
where $\otimes n$ denotes the $n-times$ convolution, $f({\bf r})$ is the common distribution of the ${\bf r}_n$ and $w(\mu_0)$ is the distribution of $\mu_0$. Making use of the notation introduced in Section \ref{Subsection_HASO3} for the multiple action of ${\bf r}_i$: $\mu_n={\bf r}_n{\bf r}_{n-1}\ldots{\bf r}_2{\bf r}_1{\bf r}_0\mu_0$, the {\em pdf} of $\mu_n$  can be rewritten as:
\begin{equation}
f(\mu_n)=f^{\otimes n}({\bf r})*w(\mu_0).
\end{equation}
Assuming again that the distribution of $\mu_0$ is a Dirac at the north pole,
the distribution of $\mu(t)$ simplifies to:
\begin{equation}
p(\mu(t)) =  \displaystyle{\sum_{n=0}^{\infty}\frac{(\lambda t)^n}{n!}e^{-\lambda t}f^{\otimes n}({\bf r})}.
\end{equation}
The assumption about $\mu_0$ is not very strong since in many practical cases the direction of propagation of the incoming wave is either known or controlled. For an isotropic random medium, we recall that the {\em pdf}s of ${\bf r}_i$ can be considered {\em zonal}. Therefore, the  
{\em pdf} $f({\bf r})$ can be expanded in a Legendre series with coefficients denoted by $\hat{f}_k$.  Using eq. (\ref{dist_RandScatt_N}) 
the {\em pdf} of $\mu(t)$ takes the form:
\begin{equation}
\begin{array}{rcl}
p(\mu(t))  & = & \displaystyle{\sum_{n=0}^{\infty}\frac{(\lambda t)^n}{n!}e^{-\lambda t}\sum_{l \geq 0}(2l+1)\left(\hat{f}_l\right)^n P_l(\mu)} \\
 & & \\
& = & \displaystyle{e^{-\lambda t}\sum_{l \geq 0}(2l+1)e^{\lambda t \hat{f}_l}P_l(\mu)}   
\end{array}  .
\end{equation}
This provides an expression of the distribution of $\mu(t)$ of the form:
\begin{equation}
p(\mu(t))=\displaystyle{\sum_{l \geq 0} (2l+1)e^{\lambda t(\hat{f}_l-1)} P_l(\mu)}.
\label{cpp_leg_exp}
\end{equation}
 Expanding   $p(\mu(t))$ in a Legendre series:
\begin{equation}
p(\mu(t))=\displaystyle{\sum_{l \geq 0}(2l+1)\hat{\mu}_lP_l(\mu)},
\label{cpp_leg_mut}
\end{equation}
a term-by-term identification in Equations (\ref{cpp_leg_exp}) and (\ref{cpp_leg_mut}) yields:
\begin{equation}
\hat{\mu}_l= e^{\lambda t (\hat{f}_l-1)}.
\label{cpp_leg_coeff_relation} 
\end{equation}
The result (\ref{cpp_leg_coeff_relation})  relates directly the Legendre coefficients $\hat{\mu}_l$ of the {\em pdf} of the scattering process $\mu(t)$ to the Legendre coefficients $\hat{f}_l$ of the {\em pdf} of the random rotations caused by the scatterers. This close link between the {\em Fourier} coefficients of these {\em pdf}s is the cornerstone of our approach as it enables us to develop a statistical estimator of the {\em Fourier} coefficients $\hat{f}_l$,
 from the observations of the {\em pdf} of $\mu(t)$.  

\section{Validity of the CPP model}\label{Section_val}

\begin{figure}
\centerline{\includegraphics[width=\linewidth]{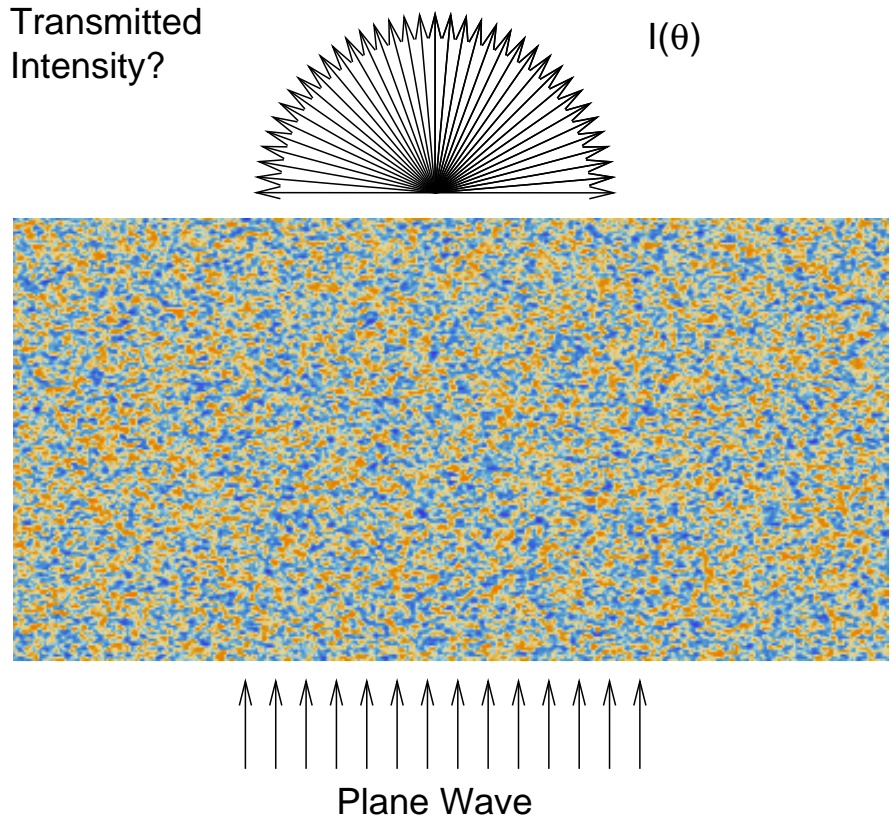}}
\caption{Position of the problem. A plane wave or a narrow beam are normally incident on a slab of random
 heterogeneous  material with thickness $z$ expressed in transport mean free path units.
 The transmitted intensity probability distribution is $I(\theta)$. }
\label{slab}
\end{figure}
\subsection{Models of random media}
In order to show the applicability of our statistical approach, we consider the following experiment 
(see Figure \ref{slab}): A plane wave or narrow beam is normally incident on a slab of random material 
described by Von Karman or Gaussian spectra, which are commonly used to describe geophysical media \cite{klimes2002}.
The angular pattern of  intensity transmitted through the slab is measured
on output. In the slab geometry with strong forward scattering, 
we expect that most of the
energy will be collected in the forward direction and therefore the CPP should
give some useful predictions of the pattern of transmitted intensity. In fact, this 
idea is confirmed by the agreement between the CPP formula (\ref{cpp_leg_exp}) and small-angle
scattering approximations of the radiative transfer equation \cite{ishimaru1978}.
It is to be noted that formula (\ref{cpp_leg_exp}) was obtained under the hypothesis that we are
able to collect all the energy at a given time $t$ but is totally free from
 a small-angle scattering assumption. 

In the CPP approach, the building block
for multiple scattering is the {\em pdf} of the random rotations that 
scatter the initial direction of the incoming beam. In the language of scattering
theory, the function that describes the angular scattering pattern is called
 the phase function and will be denoted by $f(\mu)$, where $\mu = cos(\theta)$ is the cosine of the scattering angle.
For sufficiently weak fluctuations, the phase function can be obtained from the Born approximation \cite{rytov1989} and
depends of the power spectrum of heterogeneities $\Phi$ as follows:
\begin{equation}
f(\mu) = c \Phi\left(2k\sin({\theta/2})\right),
\end{equation}
where $k$ denotes the central wavenumber of the probing wave, and $c$ is a normalization constant.
For the sake of clarity, we review basic results on  random media of special interest.
\subsubsection{Gaussian Media}
The phase function of a Gaussian random medium  is defined by the following formula:
\begin{equation}
f(\mu)=\frac{k^2a^2}{2(1 -  e^{-k^2a^2})} e^{-k^2a^2\frac{(1 - \mu)}{2}},
\end{equation}
where  $a$ is the correlation length of heterogeneities. The anisotropy parameter or mean cosine of the scattering angle is defined as: 
 \begin{equation}
 g = \int\limits^{1}_{-1} f(\mu) \mu d\mu.
\end{equation}
This parameter plays a crucial role in the definition of the transport mean free path 
 $l^* = l/(1-g)$. The transport mean free path is the typical length scale beyond which
multiple scattering can be described by a diffusion equation. In the common case of strong
forward scattering it can be much larger than the mean free path. As will be shown below,
 our approach is useful in this regime.
For Gaussian random media, one obtains $g= \coth \left(\frac{k^2a^2}{2}\right)-\frac{2}{k^2a^2}$.
The Legendre coefficients of the phase function are given by:
\begin{equation}
\hat{f}_l = \int f(\mu) P_l(\mu) d\mu.
\end{equation}
In the high-frequency limit  (large $ka$), the Legendre coefficients of the Gaussian phase function
can be approximated as:
\begin{equation}
 \hat{f}_l \approx g^{l(l+1)/2},
\end{equation}
with $g \approx 1-1/2k^2a^2$.
\subsubsection{Von Karman Media}
The Von Karman spectrum implies the following angular dependence
of the phase function:
\begin{equation}
f(\mu) = \frac{2 k^2a^2 (\alpha-1) }{\left(1-\left(4 k^2a^2+1\right)^{1-\alpha}\right)\left(1-2 k^2a^2
   (\mu-1)\right)^{\alpha}},
\end{equation}
where $\alpha$ is an exponent which controls the small-scale roughness of the medium.
Note that the phase function is normalized ($\int\limits_{-1}^{1} f(\mu) d\mu = 1$). The special cases $\alpha=3/2,2$ correspond to the well-known Henyey-Greenstein and exponential phase functions, respectively.

Using integration by parts, the coefficients of the expansion can be expressed as:
\begin{equation}
\label{bypart}
 \hat{f}_l = \frac{1}{2^n n!} \int\limits^{1}_{-1} (1- \mu^2)^l f^{(l)}(\mu) d \mu,
\end{equation}
where $f^{(l)}(\mu) $ denotes the $l^{th}$ derivative of the phase function.
Using tables of integrals, (\ref{bypart})  yields the following compact form for the Legendre coefficients 
of the Von Karman correlation function:
\begin{equation}
\begin{split}
  \hat{f}_l= & \frac{\displaystyle 2 \sqrt{\pi} (\alpha-1) (1+2ka^2)^{-(l+\alpha)} \Gamma(l+\alpha)}{\displaystyle 
\Gamma\left(\frac{3}{2}+l\right)\Gamma(\alpha)(1-(1+4ka^2)^{1-\alpha})} \\
     & \times ~{}_2F_1\left(\frac{l+\alpha}{2},\frac{1+l+\alpha}{2};\frac{3}{2}+l;\frac{4 ka^4}{(1+2ka^2)^2}\right)
\end{split} .
\end{equation}
where the definition of the hypergeometric  function $~{}_2F_1$ can be found in \cite{temme1996}.
As noted before, in the case $\alpha=3/2$, we recover the Henyey-Greenstein (HG) phase function 
which is a classically used approximation to the Mie theory for spherical scatterers.
 The Legendre coefficients can be put in the form
$\hat{f}_l = g^l$, with  $g = (1+ 2k^2a^2- \sqrt{1 + 4k^2a^2})/2 k^2a^2 $, {\em i.e.} the
Legendre coefficients are simply powers of the anisotropy parameter, valid
for any $g \in [0,1[$.
 A remarkable property of HG random media is that the convolution on the sphere of unit directions of 
$n$ HG phase functions with same parameter $g$, is a itself a HG  phase function with parameter $g^n$:
\begin{equation}
f_{HG}^{\otimes n}(\mu)=\sum_{l=0}^{\infty}(2l+1)g^{ln}P_l(\mu).
\end{equation}
\subsection{Numerical results}
\begin{figure}
\centerline{\includegraphics[width=0.9\linewidth]{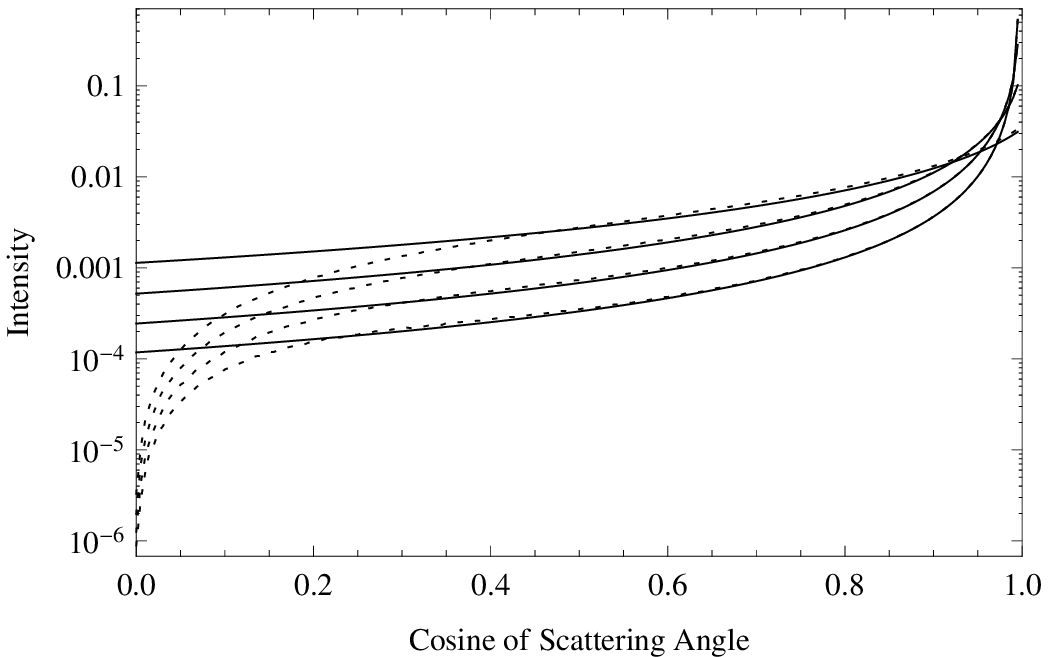}}
\caption{Comparison between Monte Carlo simulations (dots) and CPP model for the intensity
 transmitted through a slab of random medium with Heynyey-Greenstein phase function.
The four curves correspond to increasing slab thicknes $H=25/8,25/4,25/2,25 \ell$, with
$\ell$ the mean free path. The agreement is very good in a cone around the forward direction
and degrades as the scattering angle approaches grazing angles.}
\label{henyey}
\end{figure}
In order to appreciate the strengths and weaknesses of our statistical model 
 we show direct comparisons between the CPP and the radiative transfer equation for 
a Henyey-Greenstein random medium in the strong forward-scattering regime. 
 The anisotropy factor is $g=0.98$ and the slab thickness takes the
values $H=25/8,25/4,25/2,25\ell $, where $\ell$ is the mean free path of
the waves in the random medium (the transport mean free path $\ell^*$ equals $50\ell$).   For simplicity, we assume that 
the index of the slab matches that of the surrounding medium. 
The equation of radiative transfer is solved using the Monte Carlo code developed by \cite{margerin2006}
in the framework of statistically anisotropic media. 
Figure (\ref{henyey}) shows that the CPP model gives a non-uniform approximation 
 to the exact distribution of intensity transmitted through the slab 
of random material.  It is noticeable that the CPP model is always wrong in
a cone of directions perpendicular to the direction of incidence of the wave. The width of the cone
of direction increases with the slab thickness $H$. When $H$ becomes of the order
of the transport mean free path, typically, the CPP model fails to give reliable predictions
of the angular distribution of intensity. One obvious reason is that  in the CPP model, we do not prescribe any boundary condition whereas
in the slab geometry, the intensity must of course vanish in the lower hemisphere of
propagation directions. 

\begin{figure}
\centerline{\includegraphics[width=0.9\linewidth]{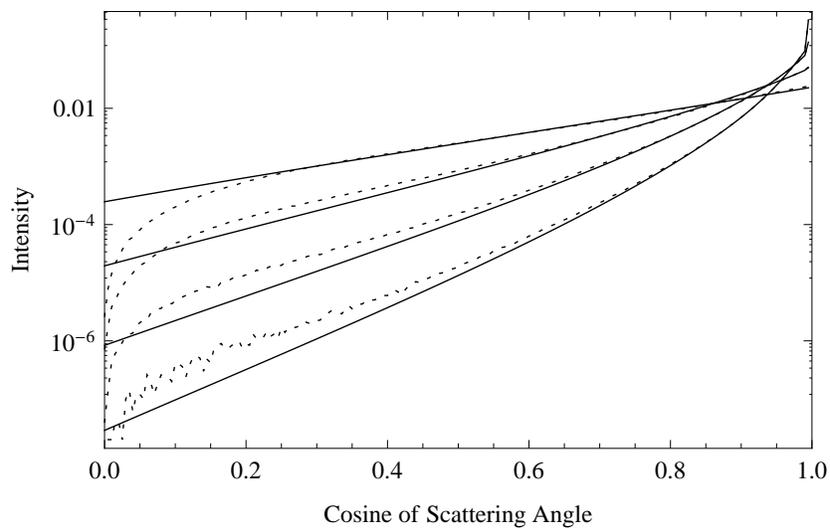}}
\caption{Comparison between Monte Carlo simulations (dots) and CPP model for the intensity
 transmitted through a slab of random medium with gaussian phase function.
The four curves correspond to increasing slab thicknes $H=25/16,25/8,25/4,25/2 \ell$, with
$\ell$ the mean free path. The agreement is very good in a cone around the forward direction
and degrades as the scattering angle approaches grazing angles.}
\label{gauss}
\end{figure}

\begin{figure}
\centerline{\includegraphics[width=0.9\linewidth]{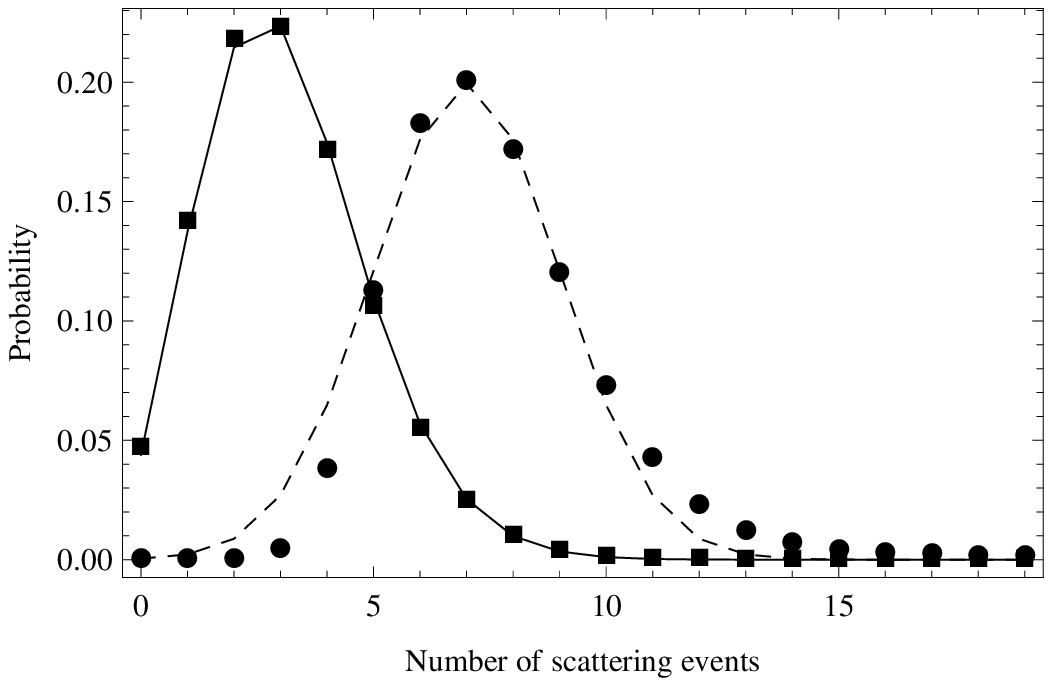}}
\caption{Distribution of the number of scattering events for the
 intensity transmitted through a slab of gaussian random medium with
thickness $H=3.125\ell$. Squares: ($0.8<\mu<1$); dots: ($0.3<\mu<0.5$).
Solid line: Poisson distribution with parameter $\lambda=3.125$.
Dashed line: Gaussian ditribution with mean $\mu=7$ and standard
 deviation $\sigma=2$.
 }
\label{probaeventg}
\end{figure}

We further remark that in the slab geometry
the distribution of scattering events does not follow a simple Poisson law, which also
limits the validity of our model.
To illustrate this point, we  examine the validity of the model in Figure \ref{gauss},  where
we show the comparison between the CPP and radiative
transfer theory for a Gaussian medium with anisotropy parameter $g=0.98$. The 
slab thickness takes the values $H=25/16,25/8,25/4,25/2\ell$. The distribution
of transmitted intensity differs markedly from the H-G case (see Figure \ref{henyey} for comparison).
 Because the Gaussian
medium is extremely smooth, the single scattering pattern is concentrated in
 a small cone of directions and large-angle scattering is excluded.
Propagation at large angle can only occur for sufficiently high-order
multiple-scattering. This entails a significant deviation of  the statistics of the number
of scattering events  from the simple Poisson distribution. This point is illustrated 
in Figure \ref{probaeventg}, where we plot the distribution of scattering events
around the forward direction ($0.8<\mu<1$) and at large angles ($0.3<\mu<0.5$) for
particles propagating through a slab of thickness $H=25/8 \ell$. While in the forward
direction the agreement between the calculated distribution of number of
 scattering events and Poisson statistics is excellent, at large
angles a clear discrepancy is observed. The distribution is  biased
towards larger number of scattering events and we verified empirically that it is in fact much better fitted by a normal distribution.
The standard deviation and mean of the Gaussian distribution have been adjusted by trial and error. The main point is to illustrate the large deviation
from the Poisson distribution. 
Note however that the Poisson law will always be well approximated by a normal
distribution (with equal mean and variance) if the number of scattering events becomes very large.

\begin{figure}
\centerline{\includegraphics[width=0.9\linewidth]{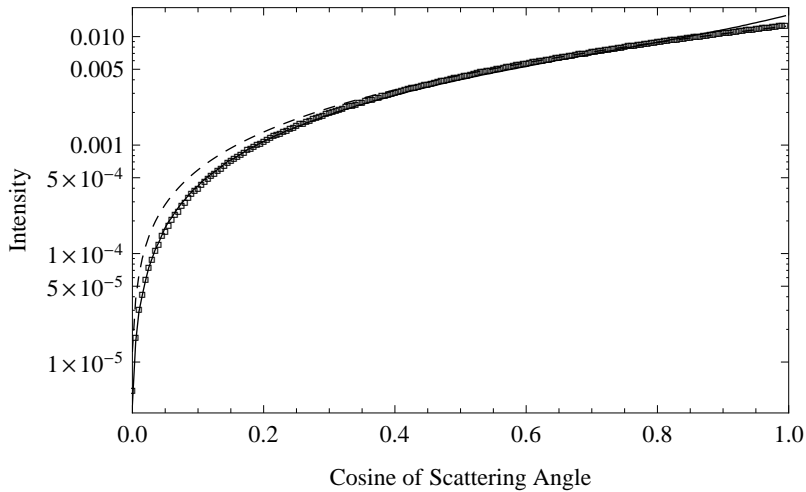}}
\caption{Transmitted intensity profile through a slab of thickness $H = \ell^*/2$ and
 anisotropy parameter $g=0.98$. Squares: Gaussian random medium. Solid line: H-G random medium.
 Dashed line: diffusion approximation. The difference between the Gaussian and H-G medium 
is only noticeable close to the forward direction ($\mu=1$).  }
\label{diffusion}
\end{figure}

We now discuss briefly the transition towards the diffusive regime.
The case of strongly anisotropic scattering has been treated analytically
in \cite{amic1996}, where it is shown that the pattern of transmitted
intensity is almost universal, independent of the type of scatterers.
In Figure \ref{diffusion}, we show the profile of transmitted intensity
obtained with the radiative tranfer approach with the slab thickness
$H=\ell^*/2$ for the Gaussian and H-G phase functions with the same
anisotropy parameter $g=0.98$. The curves for the two different
phase functions are hardly distinguishable. For comparison, we
have also plotted the simple outcome of the diffusion approximation 
$I(\mu) = \mu(1/2 + 3\mu/2)$.  
Our numerical study confirms the general conclusions of \cite{amic1996} and also demonstrates
that the diffusive regime sets in for a slab thickness of the order
of $\ell^*$ but not more.
In spite of all its deficiencies,
the CPP model gives a simple, almost analytical solution of the multiple scattering
problem and explains accurately the angular distribution of most of the forward-scattered energy.

\section{Estimation: the inverse problem}
\label{Section_estimation}


In this section, we consider the problem of estimating the Legendre coefficients of the phase function $f$ of the scatterers from the measurement of the output angular distribution $p(\mu(T))$ at a given depth in the random medium. Using the CPP model, we can construct an estimator for the Legendre coefficients of the phase function $\hat{f}_l$, using a ``decompounding'' formula. Note that the mean free time $\tau$ or equivalently (thanks to velocity normalization) the mean free path $\ell$ has to be known in the decompounding approach. Techniques to estimate the mean free path of waves in transmission geometry are described for instance in \cite{cowan1998}.
The estimators we derive below are extensions of the work done in \cite{Gug07,Buc03} for real valued random variables, to the specific case of random variables taking values over the rotation group and having {\em zonal} {\em pdf}s.    

\subsection{Estimator for the Legendre coefficients}
After proper normalization,  the transmitted intensity profile can be  interpreted as a {\em pdf} $p(\mu(T))$, where $T$ is the time required for
ballistic waves to propagate through the slab.  Let us denote by $\overline{p}(\mu(T))$ the {\em sample} {\em pdf} which corresponds to the data. The sampling resolution of $\overline{p}(\mu(T))$ depends on the acquisition system. Typically, a detector has finite aperture and therefore averages the intensity over some finite solid angle $d \Omega$. This implies that $\overline{p}(\mu(T))$ is in fact discrete  and should be more appropriately termed {\em probability mass function} with the usual normalization $\sum_{m=1}^M\overline{p}(\mu_m(T))=1$ if $M$ samples are available.

We are interested in estimating the Legendre coefficients of the {\em pdf} $p(\mu(T))$, from the observation of $\overline{p}(\mu_m(T))$. This can be
achived in  two different ways. One can either define directly the estimator on $\overline{p}(\mu(T))$, or use $\overline{p}(\mu(T))$ to generate some realizations of $\mu(T)$ and define an estimator with these realizations. In this paper, we adopt the second approach. To do so, we use a simple linear interpolation to evaluate numerically the cumulative distribution function (cdf)  and its inverse from the data $\overline{p}(\mu(T))$. This allows us to generate as many realizations of $\mu(T)$ as desired, 
by evaluating the inverse cumulative distribution function for  a collection of  uniformly distributed random numbers in $[0,1]$. 

Let us assume that we have generated ${\rm K}$ values of $\mu(T)$ using this technique and let us denote them by $\mu_{\kappa}(T)$ with $\kappa=1,\ldots,{\rm K}$. The empirical Legendre coefficients of $p(\mu(T))$, denoted by $\widetilde{\hat{\mu}}_l$ are given by:  
\begin{equation}
\widetilde{\hat{\mu}_l}=\frac{1}{{\rm K}}\sum_{\kappa=1}^{{\rm K}}P_l(\mu_{\kappa}(T)).
\label{Est_mu_empi}
\end{equation}
This estimator is unbiased, {\em i.e.} ${\mathbb E}[\widetilde{\hat{\mu}_l}-\mu_l]=0 ~, \forall ~ l$. This means that the mean value of our estimator is the real value we are looking for, {\em i.e.} $\hat{\mu}_l$. The variance of the estimator can be expressed as:
\begin{equation}
{\mathbb E}\left[\left(\widetilde{\hat{\mu}_l}-{\mathbb E}[\widetilde{\hat{\mu}_l}]\right)^2\right]=\frac{1}{{\rm K}}\left({\mathbb E}[P^2_l(\mu)] - \hat{\mu}_l^2\right).
\end{equation}
where ${\mathbb E}[P^2_l(\mu)]=\int_{-1}^1p(\mu(T)) P^2_l(\mu) d\mu$.

\begin{figure}
\centerline{\includegraphics[width=0.9\linewidth]{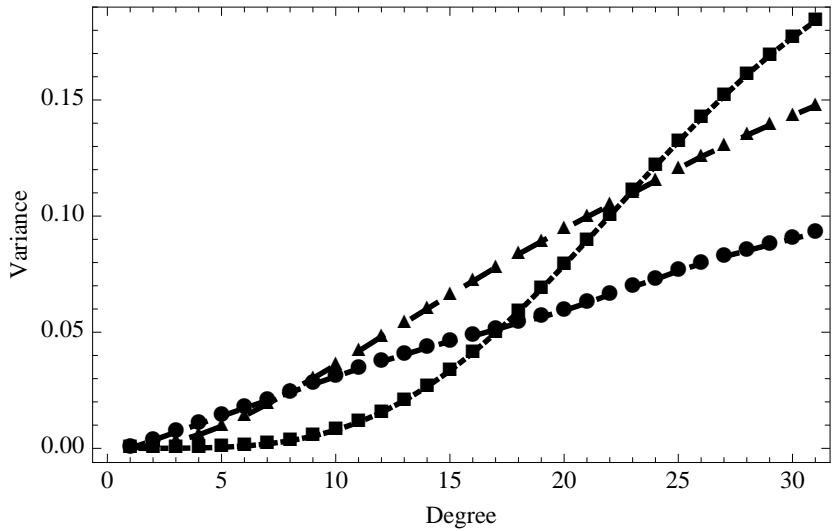}}
\caption{Variance of the Legendre coefficients empirical estimator for different phase functions: Gaussian (squares), Exponential (triangles) and Henyey-Greenstein (circles) with anisotropy parameter $g=0.995$.}
\label{varl}
\end{figure}

In the Gaussian case, the variance takes a simple analytical form, in the limit case where $(g-1) \ll 1$. Using a Taylor series expansion  of the Legendre polynomials near $\mu=1$ 
yields $Var(\widetilde{\hat{\mu}_l})=\frac{1}{K}\frac{l^2(l+1)^2}{4}(g-1)^2$.
 For other random media, such as Henyey-Greenstein or exponential, an analytical expression of the variance is not easily obtained. This is also applies to Gaussian media, away from the limiting case mentioned above. Nevertheless, the behaviour of the variance can be obtained by numerical integration. In Figure \ref{varl}, we present the variance for three different phase functions: Exponential, Gaussian and Henyey-Greenstein. The estimation has been made using $K=1000$ samples.
It can be seen on Figure \ref{varl} that the variance of the estimator increases monotonically with $l$ for the three different phase functions. While the Gaussian case follows a $4^{th}$ degree polynomial growth, Henyey-Greenstein and Exponential cases follow a more or less linear increase with the degree of the Legendre coefficients $l$. The Gaussian case is favorable for low degree coefficients while Henyey-Greenstein exhibits the lowest variance for higher degrees. Note that the variance has consequences on the ability of the proposed approach to evaluate correctly the high-degree Legendre coefficients. This will be illustrated in Section \ref{invert_sec}. 



Back to the ``decompounding'' problem, we wish to estimate the Legendre coefficients of $f$ from the samples $\mu_{\kappa}(T)$. This is possible by inverting eq. (\ref{cpp_leg_coeff_relation}) and using our empirical estimator (\ref{Est_mu_empi}). We obtain an estimate of the Legendre coefficients of the phase function of the medium:
\begin{equation}
\widetilde{\hat{f}_l}=\frac{1}{\lambda T} \ln \widetilde{\hat{\mu}_l} +1  .
\label{estfk}
\end{equation}
Equation (\ref{estfk}) is the ``decompounding'' formula. It implies that the set of Legendre coefficients $\widetilde{\hat{f}_l}$ can be estimated from the set $\widetilde{\hat{\mu}_l}$ which is directly computed from the data thanks to eq. (\ref{Est_mu_empi}). Note that it is necessary that $\widetilde{\hat{\mu}_l} > 0$. This is verified for small $l$ ($\mu_0 > \mu_l$, $\forall l\geq 1$) but may not be verified for higher degrees, mainly because of the increase of the variance of the estimator $\widetilde{\hat{\mu}_l}$ with $l$ (see Figure \ref{varl}). As a consequence, it will be necessary to truncate the number of Legendre coefficients used in the decompounding approach. The truncation degree depends on the acquisition set-up and the kind of medium investigated. This will be further illustrated in the following Section. It must be emphazised that equation (\ref{estfk}) is central as it allows  direct estimation of the heterogeneity spectrum of the medium from the output intensity distribution measurement. 


\subsection{Inverting for the random medium power spectrum}\label{invert_sec}
\begin{figure}
\centerline{\includegraphics[width=0.9\linewidth]{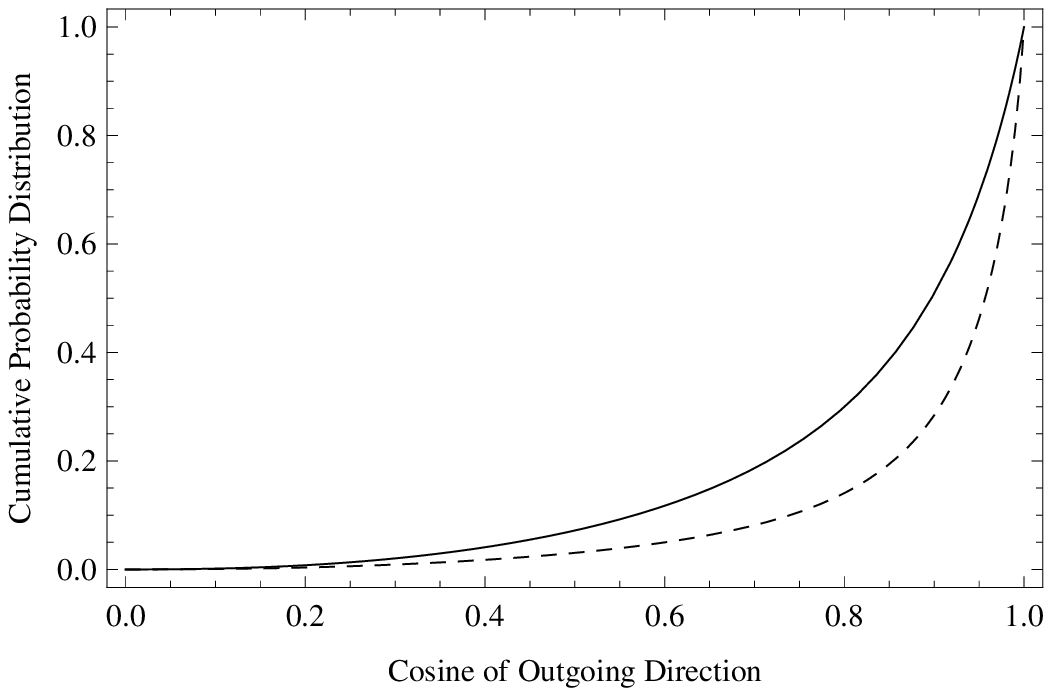}}
\centerline{\includegraphics[width=0.9\linewidth]{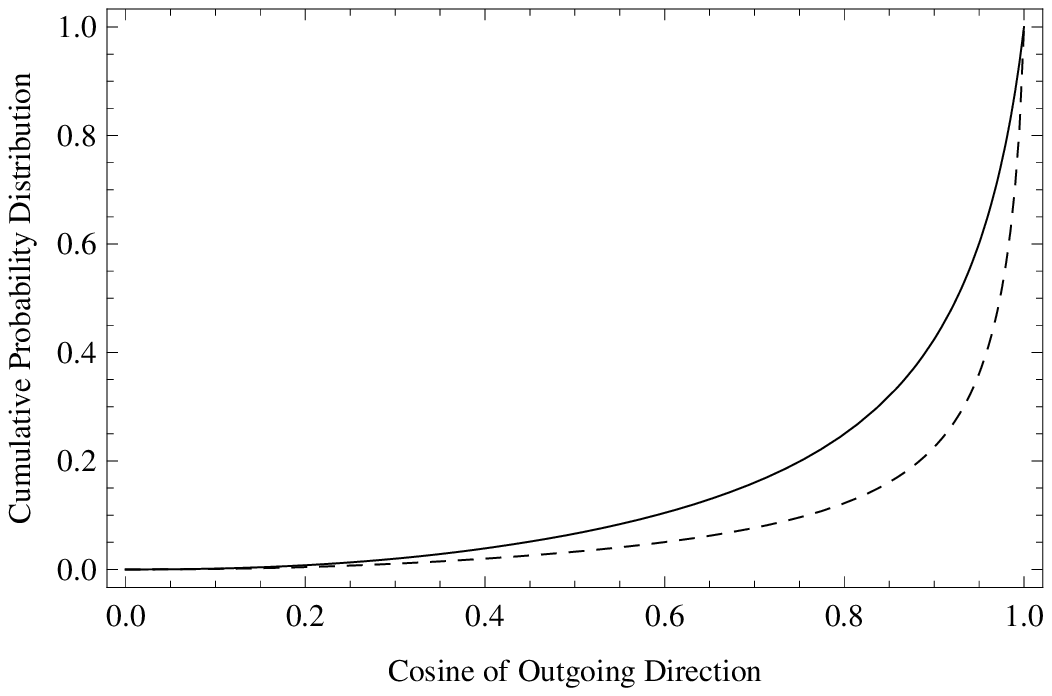}}
\caption{Cumulative distribution function of the intensity transmitted through a slab
 of random material with thickness $H=6.25 \ell$ (dashed line) $12.5 \ell$ (solid line)
 and anisotropy parameter  $g=0.98$. Top: exponential medium. Bottom: ``Kolmogorov'' medium. }
\label{cdf}
\end{figure}
\begin{figure}
\centerline{\includegraphics[width=0.9\linewidth]{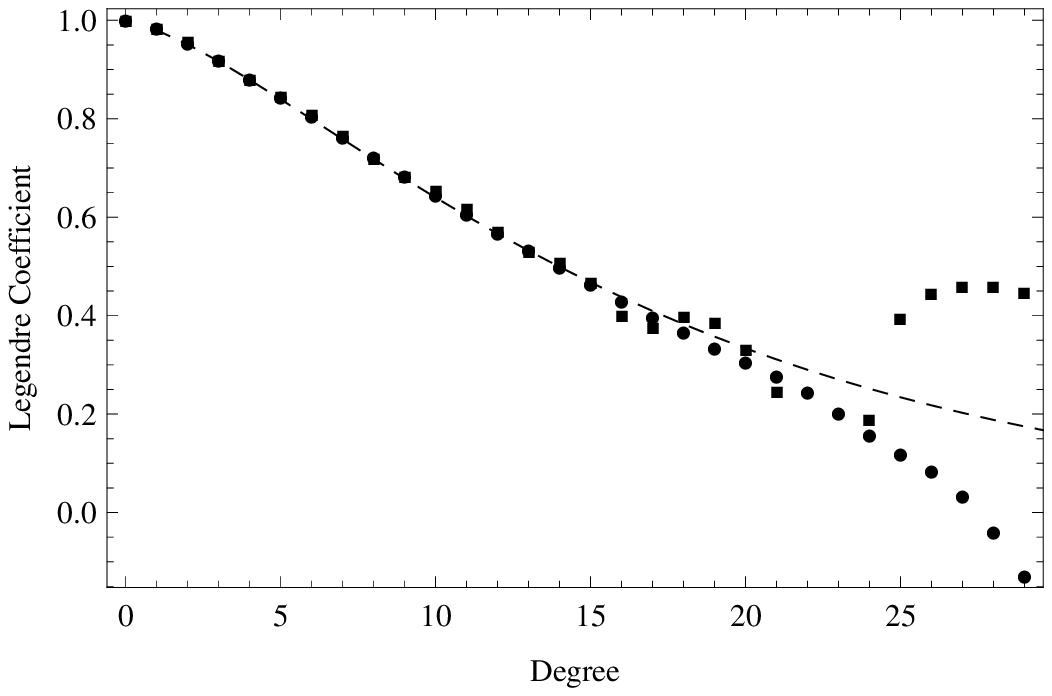}}
\centerline{\includegraphics[width=0.9\linewidth]{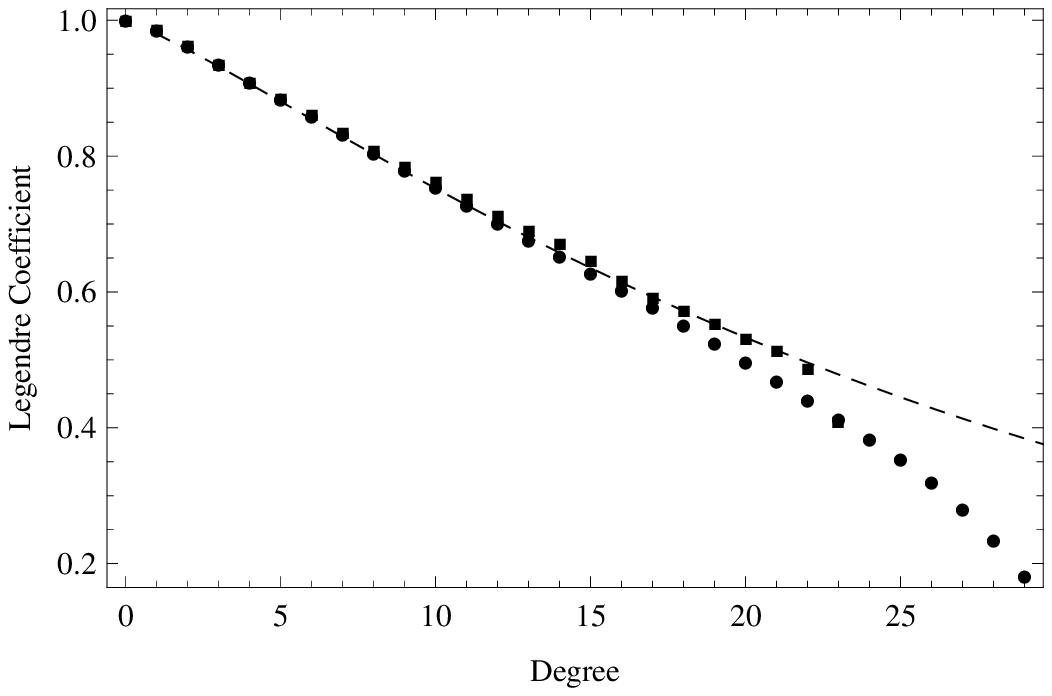}}
\caption{Estimated Legendre coefficients of the heterogeneity power spectrum for two slabs
 of random material with thickness $H=6.25 \ell$ (dots),  $12.5 \ell$ (squares)
 and anisotropy parameter  $g=0.98$. Top: exponential medium. Bottom: ``Kolmogorov'' medium. 
 The dashed line shows the exact Legendre coefficients of the distributions.}
\label{decompound}
\end{figure}
As we will now show, the CPP approach constitutes a powerful tool to invert the
power spectrum of a random medium in the regime:
 $\ell < H < \ell^* $, particularly in the regime $\ell \ll \ell^*$.
In order to demonstrate this, we consider a slab of material with Von-Karman
correlation function. The cases $\alpha=2$ (exponential correlation) and $\alpha=5/3$ (Kolmogorov
spectrum) are considered.
 Synthetic data are generated  using  Monte Carlo simulations of the scalar
radiative transfer equation in 3D.
It is to be noted that the simulations provide the intensity distribution averaged
over some finite solid angle, as will be the case in a practical experiment.
In the numerical simulations, we imposed  $d\Omega = 1/200 $ in order to obtain
sufficiently smooth results. Due to the limited averaging, the synthetic data can be noisy,
 particularly where the intensity is small but such 
phenomenon is also expected in practice. To obtain the Legendre coefficients of the intensity
distribution, we process the data as follows. From the numerical simulations,  we determine the  
cumulative distribution of the intensity. Although the original distribution can be
noisy the cumulative distribution is much smoother. Examples of empirical
$cdf$ for exponential and Kolmogorov spectra are shown in Figure \ref{cdf}.
  From the cumulative distribution,
we can evaluate the coefficients of the Legendre expansion of the intensity
by drawing a sufficiently large set of random numbers $ran$ according to:
\begin{equation}
   ran = icdf(\epsilon),
\end{equation}
where $\epsilon$ is a uniformly distributed  random number and
$icdf$ denotes the intensity inverse cumulative distribution
function obtained by linear interpolation of the original
 $cdf$.
Successive application of formula (\ref{Est_mu_empi})  and (\ref{estfk}) yields the coefficients 
$\widetilde{\hat{\mu}_l}$  and the desired Legendre coefficients $\widetilde{\hat{f}_l}$ of the
power spectrum of  heterogeneities. To obtain accurate estimates we need to draw typically $10^5$
random samples.
The comparison between the estimated coefficients and the exact Legendre expansion
for two different Von Karman random media with $g=0.98$ is shown in Figure \ref{decompound} and shows
excellent agreement. In order to check the accuracy of the decompounding formula,
it is important to analyze at least two sets of data corresponding to two different slab thicknesses.
First, this offers a consistency check for the only parameter that enters in the decompounding,
i.e. the mean free path of the waves. If there is a significant error on this quantity 
(typically more than 20 \%), the estimated Legendre coefficients of the power
spectrum will differ significantly at low degree $l$. Second, it provides a
test of validity of the inferred heterogeneity power spectrum at larger degree
$l$. Figure \ref{decompound} shows that  the estimated  $\widetilde{\hat{f}_l}$ 
for two different  slab thicknesses  split up beyond some degree $l_0$. 
For $l>l_0$, it is not possible to estimate reliably the coefficients  of the original distribution. 
It is to be noted that this test is independent of any assumption on the random medium.
\section{Conclusion}
We have developed a non parametric method to infer the properties
 of random media. Our approach relies on a generic mathematical model
 and could in principle be used to probe random  
media with acoustic, elastic, or electromagnetic waves.
The method also provides tests of  consistency
 and accuracy of the results. We show that the  angular 
distribution of intensity in a random medium
can be ``decompounded''  to estimate the  power spectrum of  heterogeneities in a random medium. An extension of the
theory to incorporate polarization measurements is currently being investigated.
\bibliographystyle{unsrt} 
\bibliography{HG_Estim_CPP}



\end{document}